\documentclass[%
reprint,
 amsmath,amssymb,
 aps,
]{revtex4-2}
\usepackage{siunitx}
\usepackage{graphicx}
\usepackage{dcolumn}
\newcommand{\RNum}[1]{\uppercase\expandafter{\romannumeral #1\relax}}
\usepackage{wrapfig}
\usepackage{float}
\usepackage{bm}
\usepackage[dvipsnames]{xcolor}

\begin{document}

\preprint{APS/123-QED}
\title{Topological Properties of SnSe/EuS and SnTe/CaTe Interfaces}

\author{Shuyang Yang}
\author{Chunzhi Wu}%
\author{Noa Marom}
\affiliation{%
 Department of Materials Science and Engineering, Carnegie Mellon University, Pittsburgh, PA 15213, USA
}%

\date{\today}

\begin{abstract}
We use density functional theory calculations to study the electronic structure of epitaxial (111) interfaces of the topological crystalline insulators SnSe and SnTe with the magnetic insulator EuS and the non-magnetic insulator CaTe, respectively. We consider both interface slab models with a vacuum region and periodic heterostructures without vacuum. We find that gaps of 21 meV at the $\Gamma$ point and 9 meV at the M point arise in the topological state at the SnSe/EuS interface, due to the magnetic proximity effect, which breaks the time reversal symmetry. The surface state at $\Gamma$ is shifted below the Fermi level by 88 meV and the surface state at M is shifted above the Fermi level by 47 meV, owing to band bending at the interface. By comparison, the topological state at the interface of SnTe/CaTe is unperturbed by the presence of non-magnetic CaTe.
\end{abstract}

\maketitle


\section{\label{sec:level1}INTRODUCTION}

Topological insulators (TIs) have attracted increasing attention for potential applications in spintronic devices \cite{wang2018spintronic,macdonald2019topological,Yu61} and quantum computing \cite{Stern1179}. Unlike ordinary insulators, although TIs are insulating in the interior, they have conducting Dirac surface states that are topologically protected by time reversal symmetry. Breaking the time reversal symmetry in TIs produces a gapped surface state, which leads to various interesting quantum phenomena, such as the topological magnetoelectric effect \cite{PhysRevLett.102.146805} and the quantum anomalous Hall effect (QAHE) \cite{PhysRevB.74.085308,PhysRevB.78.195424}. One way to break time reversal symmetry is to interface TIs with ferromagnetic insulators (FMI) \cite{Qi1184,Fujita_2011} with a magnetic easy axis perpendicular to the surface plane. The advantage of using FMIs, as opposed to metallic ferromagnets, is that an insulator does not interfere with the topological states of the TI at the interface. Several TI/FMI interfaces have been demonstrated experimentally or predicted computationally to have proximity induced gapped Dirac states, including Bi$_{2}$Se$_{3}$/MnSe \cite{PhysRevB.88.144430,PhysRevB.87.085431}, Bi$_{2}$Se$_{3}$/EuS \cite{lee2014magnetic,10.1103/physrevlett.119.027201,10.1038/nature17635}, and (Bi$_{x}$Sb$_{1-x}$)$_{2}$Se$_{3}$/yttrium iron garnet  \cite{doi:10.1063/1.4943061}.

Topological crystalline insulators (TCIs) \cite{PhysRevLett.106.106802} are a class of topological materials, whose Dirac surface states are protected not only by time reversal symmetry, but also by crystal symmetry \cite{slager2013space, PhysRevLett.106.106802,RevModPhys.83.1057}. It has been shown that some rocksalt chalcogenides with narrow inverted band gaps, such as SnTe, SnS, SnSe and Pb$_{1-x}$Sn$_x$Te(Se) are TCIs \cite{PhysRevB.88.235122,PhysRevB.89.165424,Qi1184,PhysRevB.88.045305}. For these materials, a non-zero integer mirror Chern number, which arises from band inversion at the $L$ point of the bulk Brillouin zone, ensures that the topological surface state will appear on surfaces of the (110), (111), and (001) families, preserving the mirror symmetry with respect to the \{110\} planes \cite{ando2015topological,PhysRevB.78.045426}. Experimentally, only the $\Gamma$ and $M$ Dirac state at the (111) surfaces of SnSe \cite{PhysRevX.7.041020}, SnTe \cite{PhysRevB.89.165424}, and Pb$_{1-x}$Sn$_x$Se \cite{PhysRevB.89.075317}  have been observed in angle-resolved photoemission spectroscopy (ARPES) measurements.

TCIs share many novel properties with
TIs \cite{zhang2009topological,qi2011topological}. Moreover, TCIs are predicted to exhibit QAHE with multiple dissipationless edge channels, which could offer a route for enhancing the electrical transport properties and significantly reducing the contact resistance of circuit interconnects \cite{PhysRevB.91.201401, PhysRevLett.112.046801,PhysRevLett.111.136801}. Thanks to these attractive properties, TCI/FMI interfaces may be promising platforms for the next generation of spintronic devices. Hence, it is of vital importance to explore the types of TCI/FMI interfaces that could potentially be utilized, in particular with respect to the behavior of the topological states. To date, few studies of TCI/FMI interfaces have been conducted. One experimental study of the SnTe(100)/EuS(100) interface has been reported \cite{PhysRevB.91.195310}, where the magnetic easy axis is not perpendicular but parallel to the surface plane. Therefore, the potential of TCI/FMI interfaces has not been thoroughly explored. 

We present a computational study of the TCI/FMI interface SnSe(111)/EuS(111), using density functional theory (DFT). This interface has been chosen because it is an epitaxial interface with less than 1\% lattice mismatch. In addition, EuS possesses a large magnetic moment of 7 $\mu$B \cite{mcguire1962ferromagnetism}, owing to its half filled 4f orbitals, which may potentially lead to a strong magnetic proximity effect. 
The results are compared with the TCI/ non-magnetic insulator interface, SnTe(111)/CaTe(111). The main advantage of the interfaces considered here over interfaces between Bi$_{2}$Se$_{3}$-type layered TIs with magnetic or non-magnetic insulators \cite{PhysRevB.88.144430, men2015band}, is the absence of trivial interface states. There are several possible surface terminations for SnTe(111) and the Te-terminated surface is stable under certain conditions \cite{wang2014structural, 10.1002/cphc.201300265}. For the surface of SnSe, detailed experimental investigations  have not been performed. For the purpose of comparison to SnTe(111), we focus only on the Se-terminated SnSe(111) surface. Both surface slab models (with a vacuum region) and periodic heterostructures are considered. We note that diffusion at the interface \cite{10.1021/acs.nanolett.7b00560, 10.1021/acs.nanolett.8b03057} is not considered here. We find that a gapped surface state appears at the SnSe(111)/EuS(111) interface. In contrast, the topological state at the SnTe(111)/CaTe(111) interface remains unperturbed. Therefore, the opening of a gap at the interface may be attributed to the presence of EuS. 


\section{\label{sec:level1}METHODS}

DFT calculations were performed using the Vienna ab initio simulation package (VASP) \cite{PhysRevB.47.558} with the projector augmented wave method (PAW) \cite{PhysRevB.50.17953,PhysRevB.59.1758}. The generalized gradient approximation (GGA) of Perdew, Burke, and Ernzerhof (PBE) was employed for the description of exchange-correlation interactions among electrons \cite{PhysRevLett.77.3865,PhysRevLett.78.1396}. A Hubbard U correction was used for the \textit{f} orbitals of Eu within the Dudarev approach \cite{PhysRevB.57.1505}. The value of U - J was found by the linear response method \cite{PhysRevB.71.035105} to be 7.6 eV, as detailed below. Spin-orbit coupling (SOC) \cite{PhysRevB.93.224425} was used. Key tags for convergence were BMIX=3, AMIN=0.01 and ALGO=Fast. In addition, dipole corrections \cite{PhysRevB.46.16067} were included, using the IDIPOLE=3 tag. A 9$\times$9$\times$1 k-point grid was used to sample the Brillouin zone. Structural relaxation was performed until the change of the total energy was below  10$^{-5}$ eV. Projected band structure calculations were performed using the tag LORBIT=11, which outputs the weights of states of certain atoms in the PROCAR file. In the following, we provide a complete account of the determination of U by the linear response method with VASP (Section A) and the construction of surface and interface slab models (Section B).

\subsection{\label{sec:level2}Linear response method for GGA+U}
EuS is a semiconductor with a band gap of 1.64 eV \cite{PhysRevB.9.2513, PhysRevB.70.115211}. However, the PBE functional yields no gap. To rectify this, we use the GGA+U method.  According to Ref. \cite{PhysRevB.71.035105}, the effective Hubbard U value for a particular orbital can be calculated from first principles considerations by 

\begin{equation}
U = \chi^{-1}_{0} - \chi^{-1}
\end{equation}
\begin{equation}
\chi_{0} = \frac{\partial n_{0}}{\partial \alpha} 
\end{equation}
\begin{equation}
\chi = \frac{\partial n}{\partial \alpha} 
\end{equation}

where $\chi_{0}$, $\chi$ are the linear response coefficients for non-charge-self-consistent response and charge-self-consistent response to an applied local potential, $\alpha$, respectively. $n$ and $n_{0}$ are the total number of electrons occupying the orbital in question.
To use this method in VASP, LDAUTYPE=3 should be specified in the INCAR file. Then, the LDAUU and LDAUJ parameters are regarded as the identical $\alpha$ parameters for the up and down spin channels rather than their actual physical meaning. Here, we demonstrate the application of the linear response method to the \textit{f} orbitals of Eu in EuS.

\begin{figure}
\includegraphics[width=3.2in, height=2in]{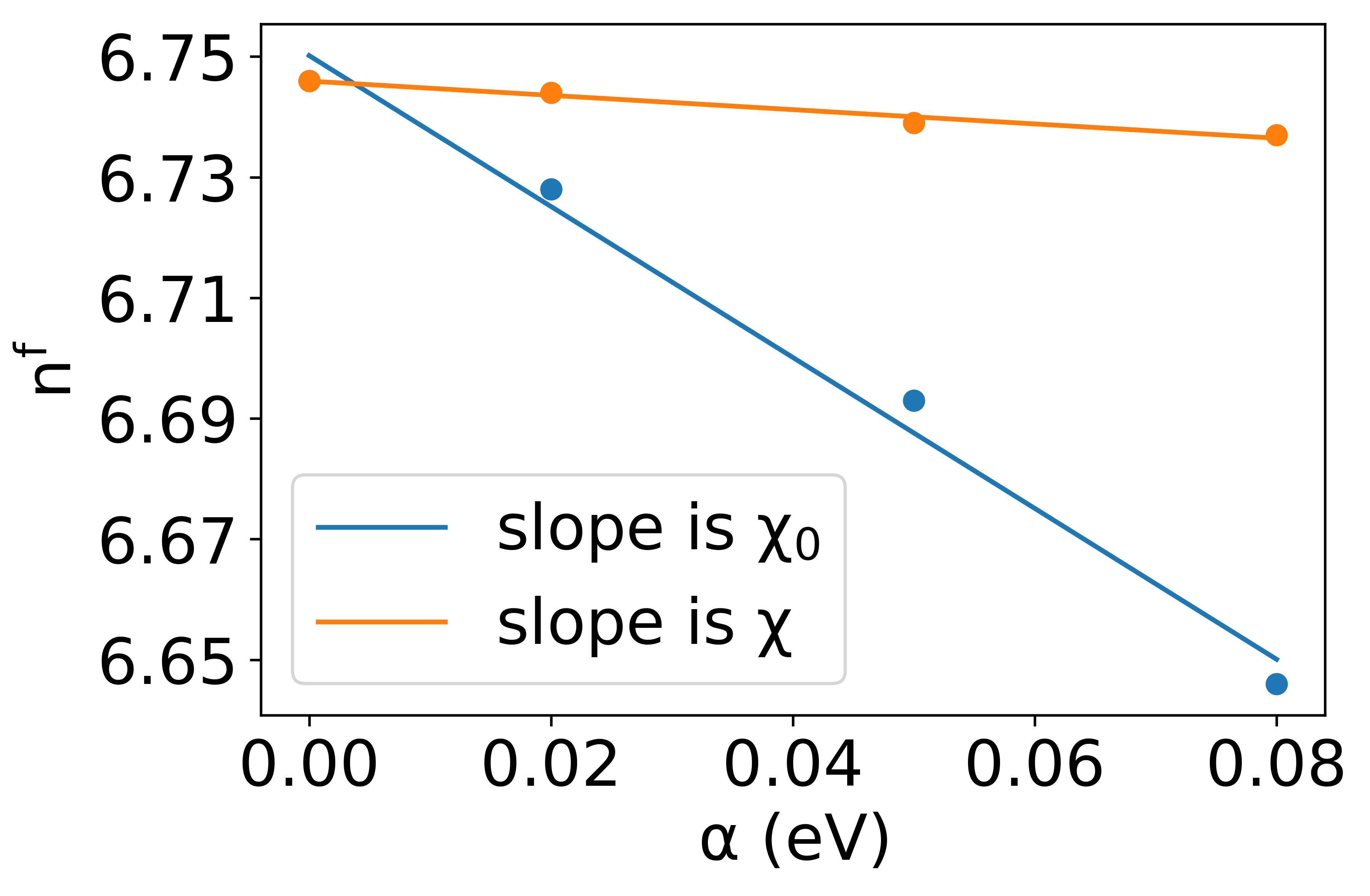}
\caption{\label{fig:lr} Occupation of the \textit{f} orbital of the Eu atom as a function of the local potential applied in a non-charge-self-consistent calculation (blue) and a charge-self-consistent calculation (orange). The non-self-consistent response $\chi_{0}$, and the self-consistent response, $\chi$, are obtained from the slope of a linear fit. The data point at $\alpha$ = 0 is shared by both lines.}
\end{figure}

We constructed a 3 $\times$ 3 $\times$ 3 super-cell for EuS. Four $\alpha$ values of 0, 0.02, 0.05, and 0.08 eV were applied for the first Eu atom in the POSCAR file, which was labeled as species "Eu Eu S" instead of "Eu S". Then, the occupations of the Eu \textit{f} orbital, $n^{f}_{0}$ and $n^{f}$, were calculated with the non-charge-self-consistent tag (ICHARG=11) and the charge-self-consistent tag (ICHARG=2). Finally, a linear fit was performed to obtain $\chi_{0}$ and $\chi$, as shown in Fig \ref{fig:lr}. Based on this, the effective U was found to be 7.6 eV, which is consistent with the value of 7.4 eV reported in \cite{10.1103/physrevb.75.045114}.  This value of U yields a band gap of 1.16 eV for EuS. We confirmed the U value is converged with respect to super-cell size by conducting additional calculations using super-cells of 1 $\times$ 1 $\times$ 1  and 2 $\times$ 2 $\times$ 2, which yielded U values of 5.9 eV and 7.4 eV, respectively.    

\subsection{\label{sec:level2}Slab construction}
We constructed six slab models: surface models for the pure TCIs with 75 layers of SnSe and SnTe and a 40 \si{\angstrom} vacuum region; interface models with 75 layers of SnSe on top of 12 layers of EuS and 41 layers of SnTe on top of 26 layers of CaTe and a 40 \si{\angstrom} vacuum region; and periodic heterostructures with 75 layers of SnSe on top of 13 layers of EuS  and 41 layers of SnTe on top of 31 layers of CaTe without vacuum. For the surface and interface models with vacuum the S, Te, and Se atoms at the surface were passivated by hydrogen atoms to eliminate topologically trivial surface states. We assumed that epitaxially matched SnSe and SnTe films would grow on top of EuS and CaTe substrates, respectively. The experimental lattice constants of the substrates were utilized for all slab models, as detailed in Table \ref{tab:table1}. At the SnSe/EuS and SnTe/CaTe interfaces, we confirmed that bonding between Se-Eu and between Te-Ca is favorable over the bonding between Se-S and Te-Te, respectively, as expected. FCC and HCP stacking configurations were considered for the interfacial position of Se/Te with respect to Eu/Ca. The FCC configuration was found to be more stable by 0.4 eV for SnSe/EuS and by 10.3 eV for SnTe/CaTe. Therefore, the FCC configuration is considered hereafter.

\begin{table}[H]
\caption{\label{tab:table1}%
Lattice constants and mismatch for SnSe, SnTe, SnSe/EuS, SnTe/CaTe.
}
\begin{ruledtabular}
\begin{tabular}{ccc}
\textrm{Structures}&
\textrm{Lattice constants(\si{\angstrom})}&
\textrm{Lattice mismatch(\%)}\\
\colrule
SnSe & 5.99\cite{nikolic1965optical} & -\\
SnTe & 6.31\cite{PhysRevB.89.165424} & -\\
SnSe/EuS\ & 5.97\cite{doi:10.1080/10408437208244865}& 0.3\\
SnTe/CaTe & 
6.35\cite{khachai2009first} & 0.6 \\
\end{tabular}
\end{ruledtabular}
\end{table}

\begin{figure}[H]
\includegraphics[width=3.5in, height=2.19in]{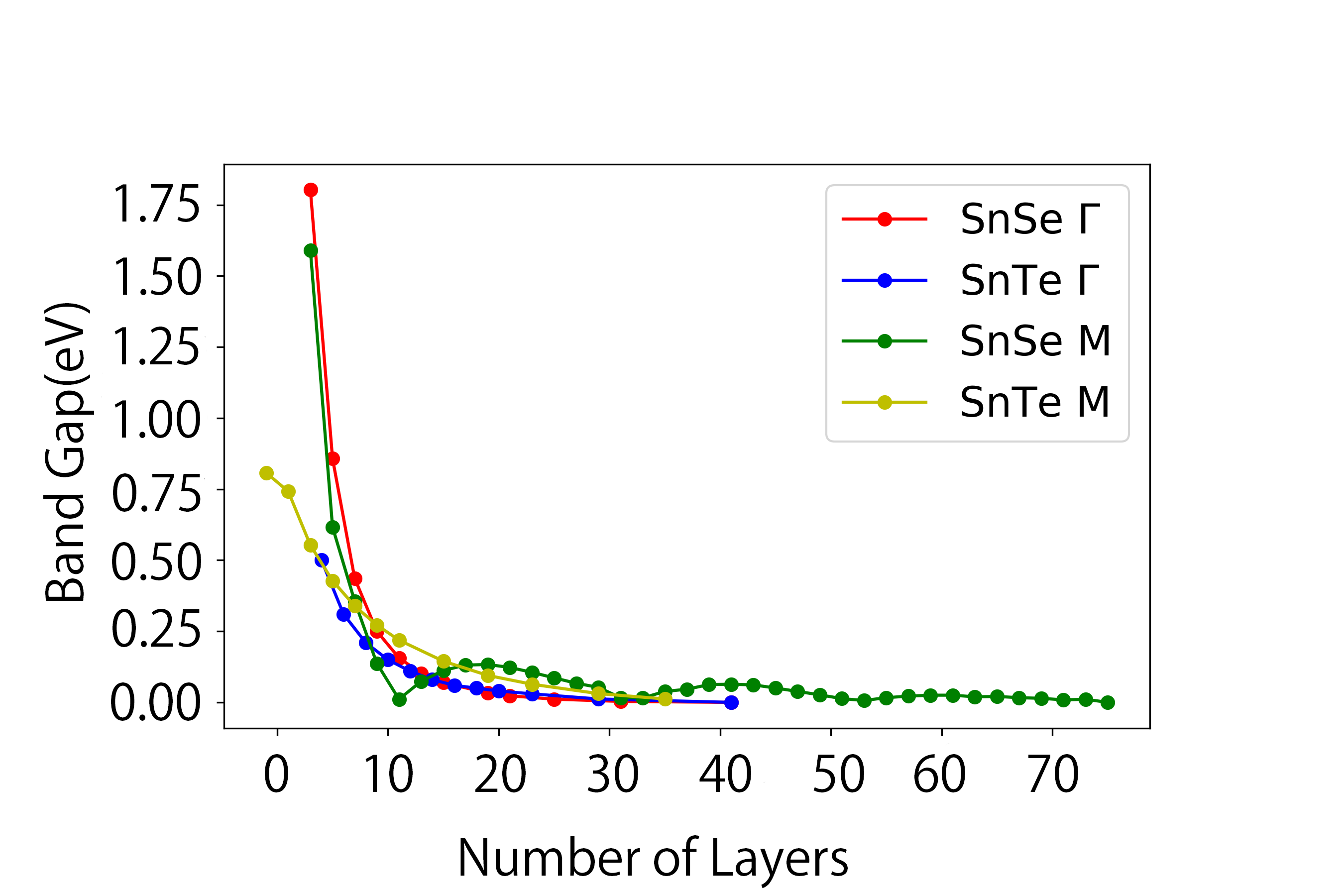}
\caption{\label{fig:thickness} The band gap at $\Gamma$ and $M$ as a function of the  number of TCI atomic layers.}
\end{figure}

When constructing interface models it is important to converge the number of layers. In particular, the Dirac state of TIs and TCIs is very sensitive to the number of layers \cite{Safaei_2015,PhysRevB.90.045309}. Figure \ref{fig:thickness} shows the band gap of SnSe/EuS and SnTe/CaTe surface models as a function of the number of atomic layers of SnSe and SnTe. For SnTe, with 41 layers the gap closes and the conducting surface state appears. While for SnSe, it requires 75 layers to close the gap at the $M$ point.

Finally, in order to determine the direction of the magnetic easy axis of EuS at the interface with SnSe, we calculated the magnetic anisotropy energy with lattice parameters of 5.97$\AA$ (corresponding to bulk EuS) and 5.99 $\AA$ (corresponding to bulk SnSe). We find that the energy of out-of-plane magnetization is lower than that of in-plane magnetization by 0.42 meV and 0.47 meV, respectively. This means that the magnetic easy axis of EuS is perpendicular to the interface with SnSe. This is in contrast to the Bi$_{2}$Se$_{3}$/EuS interface, for which the magnetic easy axis of EuS has been reported to switch from out-plane to in-plane when the lattice parameter changes from 5.97$\AA$ to 5.99 $\AA$ \cite{10.1103/physrevlett.119.027201,10.1038/nature17635}.  
 
\section{\label{sec:level1}RESULTS AND DISCUSSION}

\begin{figure*}
\includegraphics[width=7in, height=6in]{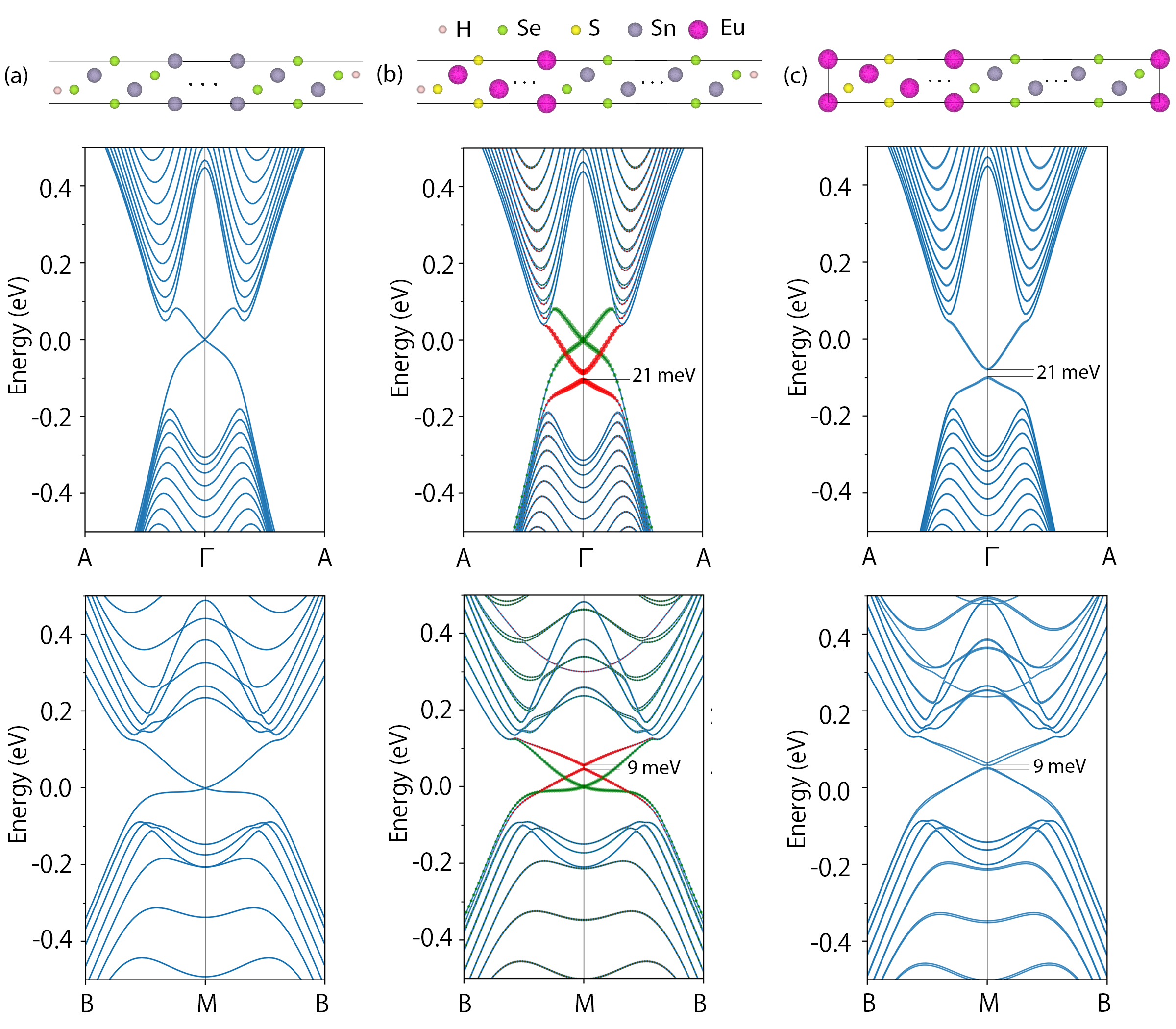}
\caption{\label{fig:snse-eus} Geometries and band structures for (a) a surface slab of SnSe(111), (b) an interface slab of SnSe(111)/EuS(111), and (c) a periodic heterostructure of SnSe(111)/EuS(111) without vacuum space. The bottom (top) of the each structure is located at the left (right) side. A and B are points along the $K - \Gamma$ and $M - K$ paths, respectively, with the coordinates (0.1, -0.1, 0) and (0.449, -0.102, 0). In panel (b), the areas of the green and red circles correspond to the weights of states originating from the top two layers of SnSe and the bottom two layers of SnSe, respectively.}
\end{figure*}


Figure \ref{fig:snse-eus}(a) shows the band structures of a surface slab of SnSe(111). The topological surface states are clearly visible at the Fermi level at the $\Gamma$ (top) and $M$ (bottom) points. Figure \ref{fig:snse-eus}(b) shows the band structures of an interface slab model of SnSe(111)/EuS(111) and Figure \ref{fig:snse-eus}(c) shows the band structures of a periodic heterostructure of SnSe(111)/EuS(111). From comparing panels (b) and (c) to panel (a), it is evident that only the states of SnSe are contributing to the band structure in the vicinity of the Fermi level. The bands of EuS are sufficiently far from the Fermi level to be completely separated from the surface Dirac cones for both interface models. This demonstrates a distinctive advantage of FMIs, such as EuS, compared to metallic ferromagnets \cite{baker2015spin,wang2017room} or highly doped substrates \cite{checkelsky2012dirac,chang2015high,haazen2012ferromagnetism}. 

At both the $\Gamma$ and $M$ points, the band structures of the interface slab in panel (b) exhibit two Dirac cones, a gapped one and a gapless one. In comparison, the SnSe slab in panel (a) and the periodic heterostructure in panel (c) show only one degenerate Dirac cone, which is gapless for pure SnSe and gapped for the periodic heterostructure. The areas of the green and red circles in panel (b) correspond to the weights of states originating from the top two layers of SnSe and the bottom two layers of SnSe, respectively. This indicates that the gapped Dirac cone is an interface state, contributed by the two layers adjacent to the EuS, whereas the gapless Dirac cone is a surface state, contributed by the two layers adjacent to the vacuum region. The opening of a gap may be attributed to proximity to EuS. A magnetic moment is induced onto SnSe, which breaks the time reversal symmetry and opens a gap of 21 meV at $\Gamma$ and 9 meV at $M$ in the topological states at the interface with EuS. 

The magnetic moment of the interfacial Eu atom is 6.980 $\mu B$, close to the magnetic moment of bulk EuS, which is 6.967 $\mu B$. The largest proximity induced magnetic moment on the Se layer adjacent to the Eu is -0.31 $\mu B$. Because the \textit{f} orbitals of Eu are highly localized the proximity induced magnetism is short-ranged and the magnitude of the induced magnetic moment decreases rapidly with  the distance from the interfacial Eu. The induced magnetic moment on the next layer of Sn is -0.004 $\mu B$. Because the magnetic proximity effect in the TCI is confined to the two layers closest to the interface and the induced magnetic moments are small, the Dirac cone at the top surface (next to the vacuum region) is not influenced and remains intact.

The binding energy for topological surface states is defined as the energy difference between the Dirac point and the Fermi level \cite{xia2009observation}. At the $\Gamma$ point, the interface state (indicated by red circles in Figure \ref{fig:snse-eus}(b)) is shifted downward from the Fermi level and has a binding energy of -88 meV, such that it is separated from the top surface state (indicated by green circles). At the $M$ the interface state is shifted upwards from the Fermi level and has a binding energy of +47 meV

\begin{figure*}
\includegraphics[width=7in, height=3.5in]{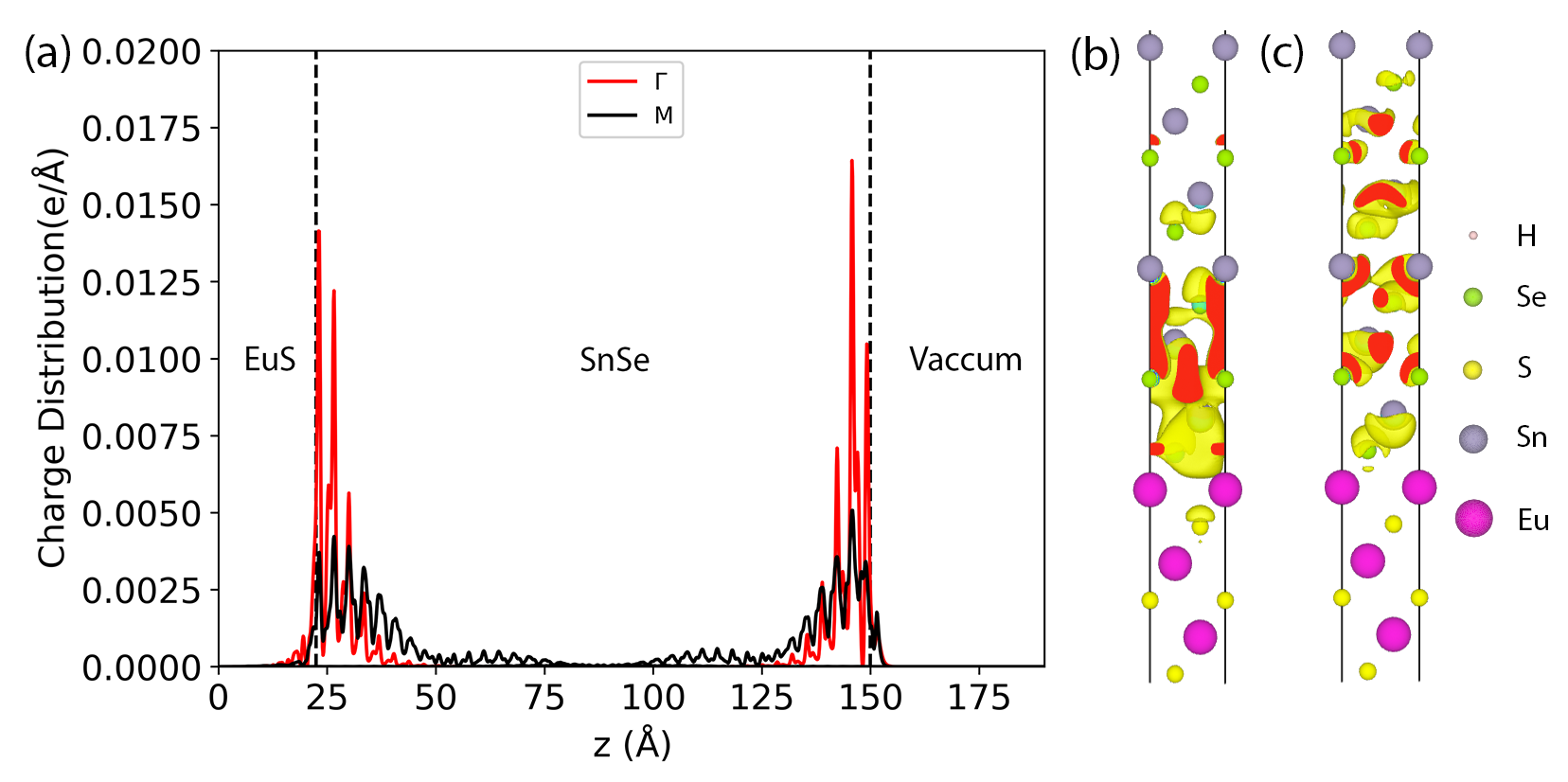} 
\caption{\label{fig:partial_chg}(a) Real space partial charge distributions of the topological interface states at points $\Gamma$ and $M$. The partial charges associated with the Dirac states at $\Gamma$ and $M$ are shown in panels (b) and (c), respectively. The electron charge iso-surface is colored in yellow and its intersections with the unit cell boundary are colored in red.}
\end{figure*}

To investigate the range of the proximity effect and elucidate the difference in the gap at the $\Gamma$ and $M$ points, we computed the real space partial charge distributions of the Dirac states at $\Gamma$ and $M$. The results are shown in Figure \ref{fig:partial_chg}. For van-der-Waals-layered TIs, the Dirac states escape the interface region, irrespective of whether the contact material is a ferromagnetic insulator \cite{PhysRevB.88.144430} or a non-magnetic insulator \cite{men2015band}. In contrast, at the interface with a non-layered TCI, such as SnSe, the topological states do not escape the near-interface layers. This leads to stronger hybridization with the magnetic EuS, which results in a much larger exchange gap than in Bi$_{2}$Se$_{3}$/EuS. The difference in spatial localization of the $\Gamma$ and $M$ Dirac states explains the difference in the gap value. The topological state at the $\Gamma$ point is concentrated closer to the interface and penetrates farther into the EuS than the topological state at the $M$ point. Therefore, the exchange gap is significantly larger at the $\Gamma$ point. Moreover, owing to the fact that the induced magnetic moment is highly localized and is mostly confined to the two atomic layers closest to the interface, the gap of the interface state converges even with just two atomic layers of EuS, and remains approximately constant regardless of the number of EuS layers. 

The periodic heterostructure of SnSe/EuS has only interfaces and no surface. Therefore, it exhibits a degenerate gapped Dirac cone at $\Gamma$. For the Dirac cones at $M$, there is a slight deviation from degeneracy,  possibly because the periodic heterostructure model with 75 layers is still under-converged. Similar to the interface state found in the slab model, the gaps at $\Gamma$ and $M$ are 21 meV and 9 meV, respectively, and the Dirac cones are shifted from the Fermi level by -88 meV and +47 meV. We note that the gapped Dirac cone does not occur if one simply changes the lattice parameter of SnSe from 5.99 to 5.97 \si{\angstrom} without the presence of EuS. The  gap of 21 meV at the $\Gamma$ point at the SnSe/EuS interface is significantly larger than the 9 meV gap reported previously for Bi$_{2}$Se$_{3}$/EuS \cite{lee2014magnetic}.

\begin{figure*}
\includegraphics[width=7in, height=6in]{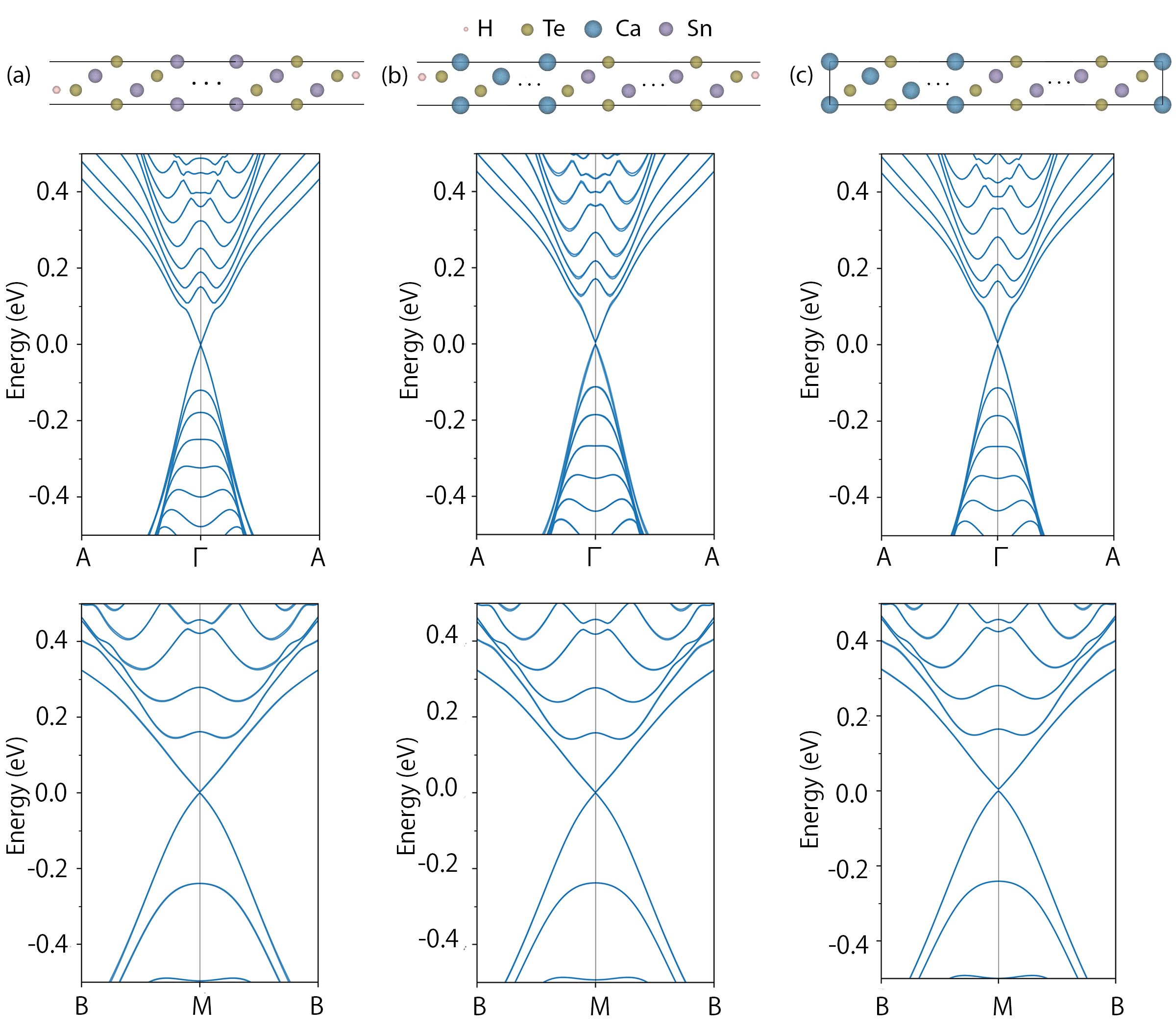}
\caption{\label{fig:snte-cate}Geometries and band structures for (a) a surface slab of SnTe(111), (b) an interface slab of SnTe(111)/CaTe(111), and (c) a periodic heterostructure of SnTe(111)/CaTe(111) without vacuum space. The bottom (top) of the each structure is located at the left (right) side. A and B are points along the $K - \Gamma$ and $M - K$ paths, respectively, with the coordinates (0.1, -0.1, 0) and (0.449, -0.102, 0).}
\end{figure*}   

\begin{table}[H]
\caption{\label{tab:bader charge}%
The amount of charge lost or gained for each atom at the SnSe/EuS interface, relative to bulk SnSe, obtained by Bader analysis. The layer index starts from the interface side.
}
\begin{ruledtabular}
\begin{tabular}{ccc}
\textrm{Layer index}&
\textrm{Charge of Sn (e)}&
\textrm{Charge of Se (e)}\\
\colrule
1    &    0.2242    &    -0.0198   \\
2    &    0.0164    &    -0.0033 \\
3    &    0.0006    &    0.0250    \\
4    &    -0.0041    &    0.0065    \\
5    &    -0.0004    &    0.0003   \\
6    &    -0.0001    &    -0.0013    \\
7    &    -0.0001    &    0.0175    \\
8    &    -0.0082    &    0.0201    \\
9    &    -0.0006    &    -0.0017    \\
10    &    0.0002    &    0.0004    \\
\end{tabular}
\end{ruledtabular}
\end{table}

\begin{table}[H]
\caption{\label{tab:bader charge1}%
The amount of charge lost or gained for each atom at the SnSe/EuS interface, relative to bulk EuS, obtained by Bader analysis. The layer index starts from the interface side.
}
\begin{ruledtabular}
\begin{tabular}{ccc}
\textrm{Layer index}&
\textrm{Charge of Eu (e)}&
\textrm{Charge of S (e)}\\
\colrule
1     &0.0429&-0.0259  \\
2    &   0.0003 &-0.0268  \\
\end{tabular}
\end{ruledtabular}
\end{table}

To further demonstrate that the gapped interface state is a result of proximity to the ferromagnetic insulator EuS, we compare the SnSe/EuS interface to the interface of the TCI SnTe with the non-magnetic insulator CaTe \cite{li2014superlattice}. Figure \ref{fig:snte-cate}(a) shows the band structure of a surface slab of SnTe(111). The topological surface states at the $\Gamma$ point (top) and $M$ point (bottom) are clearly visible at the Fermi level, similar to the SnSe(111) surface. Figure \ref{fig:snte-cate}(b) shows the band structures of an interface slab model of SnTe(111)/CaTe(111) and Figure \ref{fig:snte-cate}(c) shows the band structures of a periodic heterostructure of SnTe(111)/CaTe(111). In sharp contrast to SnSe/EuS, both the interface slab model and the periodic heterostructure of SnTe/CaTe show no significant changes. In both cases the Dirac cones at $\Gamma$ and $M$ remain unperturbed because CaTe is non-magnetic and therefore its presence does not break the time reversal symmetry. Moreover, the Dirac point remains at the Fermi level. 

To explain why the Dirac cone is shifted for the SnSe/EuS interface but not for the SnTe/CaTe interface, we examine the electrostatic potential and charge transfer. Charge transfer and the local electrostatic potential at the interface are important features because they can show if there is band bending \cite{littlejohn1983effects,feenstra2006influence,zhang2012band}.
The charge transfer at an interface is given by: 
\begin{figure*}
\includegraphics[width=7in, height=4.5in]{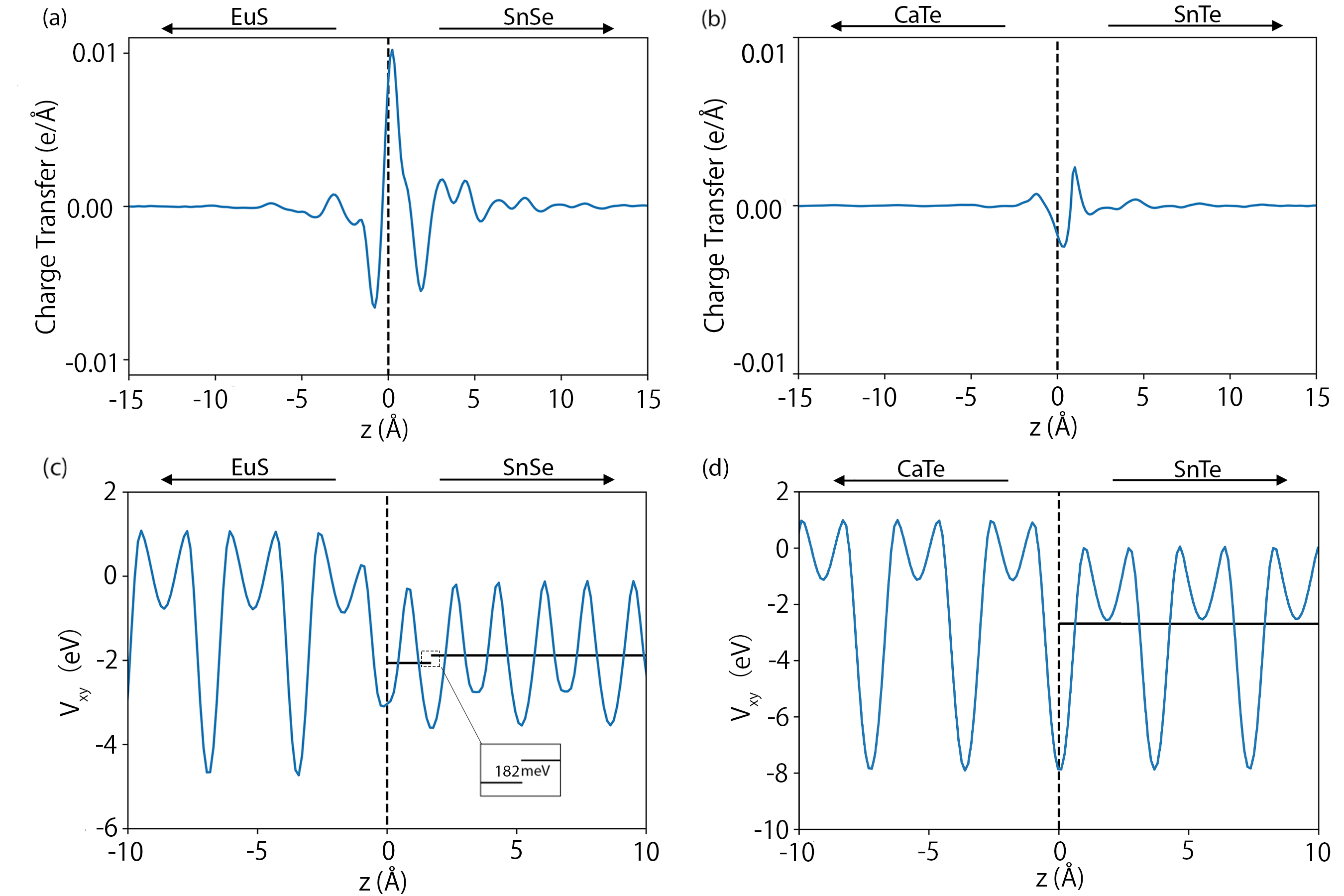} 
\caption{\label{fig:chg}Charge transfer as a function of distance from the interface for slab models of (a) SnSe/EuS and (b) SnTe/CaTe. Electrostatic potential $V_{xy}$ along z-axis for slab models of (c) SnSe/EuS and (d) SnTe/CaTe. The solid lines represent the fitted potential. The center of the interface is defined as $z = 0$.}
\end{figure*}
\begin{equation}
C_{net}(z) = C[interface] - C[substrate] - C[film]
\end{equation}

where $C[\cdot]$ is the charge averaged over the $xy$ plane along the $z$-axis. The center of the interface is defined as $z = 0$. To estimate the charge, DFT calculations were performed for the interface slab, and for separate slabs containing only the substrate and only the film with the same geometry as the interface slab. The separate film and substrate slabs were passivated with hydrogen. The results for SnSe/EuS and SnTe/CaTe are shown in panels (a) and (b) of Figure \ref{fig:chg}, respectively. In both cases, $C_{net}$ is negative at the interfacial layer of Eu(Ca) and positive at the first layer of Se(Te), meaning that charge is transferred from the EuS and CaTe substrates to the SnSe and SnTe films. To find the total charge transferred we integrated over $C_{net}(z)$ from the bottom of the slab to $z = 0$. We find that $0.005 e$ is transferred from EuS to SnSe and $0.0001 e$ is transferred from CaTe to SnTe, \textit{i.e.,} the charge transfer at the SnSe/EuS interface is considerably larger than the charge transfer at the SnTe/CaTe interface. 

The electrostatic potential along the $z$-axis, averaged over the $xy$ plane, for SnSe/EuS and SnTe/CaTe interface slab models is shown in panels (c) and (d) of Figure \ref{fig:chg}, respectively.  To investigate whether there is a jump in the potential at the interface, we averaged over the potential from $z = 0$ to the second Se/Te atomic layer and from the second Se/Te atomic layer to the top of the slab (near the vacuum region). We find that the redistribution of charge at the SnSe/EuS interface leads to a jump of 182 meV in the averaged electrostatic potential of SnSe. In contrast, the potential change at the SnTe/CaTe interface is close to zero as no discontinuity can be seen in panel (d). The charge transfer and change in the potential at the SnSe/EuS interface cause band bending and shifts the surface states of the TCI with respect to the Fermi level. For the SnTe/CaTe  interface, the charge transfer and change in the electrostatic potential are negligible. Therefore, the Dirac cone remains at the Fermi level.

To explain why the interface states at the $\Gamma$ and $M$ points shift in opposite directions, we utilize Bader charge partitioning \cite{sanville2007improved} to obtain the fractional charges for each atom. The results are shown in Table \ref{tab:bader charge} and \ref{tab:bader charge1}. We find that the first two Sn atoms gain charge while the Se atoms lose charge. In contrast, the third to tenth layers of Sn and Se atoms behave mostly in an opposite manner. Based on the partial charge distribution shown in Figure \ref{fig:partial_chg}, the interface state at the $\Gamma$ point is mostly confined to the first few layers, whereas the interface state at $M$ is distributed over more layers, deeper into the SnSe. Thus, we may attribute the opposite shifts of the interface states at the $\Gamma$ and $M$ points to the opposing direction of charge transfer at the interface vs. deeper in the SnSe. Overall, the binding energy of -88 meV and +47 meV of the gapped topological states at the SnSe/EuS interface is small compared to the binding energies of -0.8 eV and -0.234 eV reported for Bi$_{2}$Se$_{3}$/MnSe \cite{PhysRevB.88.144430,PhysRevB.87.085431} and Bi$_{2}$Se$_{3}$/EuS \cite{lee2014magnetic}, respectively. It may be possible to manipulate the position on the Dirac cones in experiments by electric gating \cite{xiu2011manipulating,checkelsky2011bulk} or doping \cite{arakane2012tunable,hsieh2009tunable,wang2010tuning}.

\section{\label{sec:level1}CONCLUSION}
In summary, we have studied the topological properties of SnSe/EuS and SnTe/CaTe (111) interfaces by first principles simulations. In the former, the TCI SnSe is in contact with the ferromagnetic insulator EuS, whereas in the latter the TCI SnTe in contact with the non-magnetic insulator, CaTe. We find that both interfaces have no trivial interface states. At the TCI/FMI interface the magnetic proximity effect breaks the time reversal symmetry and opens a gap of 21 meV at the $\Gamma$ point and a gap of 9 meV at $M$ in the topological interface state of SnSe. Charge transfer at the interface leads to band bending and shifts the gapped Dirac states below the Fermi level by 88 meV at $\Gamma$ and above the Fermi level by 47 meV at $M$. A slab model of the SnSe/SuS interface shows that the magnetic proximity effect is very localized and confined to the SnSe surface directly in contact with EuS. The interface state at the $\Gamma$ point penetrates deeper into the EuS than the interface state at $M$, leading to a larger exchange gap opening at $\Gamma$. In comparison, the topological state of SnTe is unperturbed by the presence of the non-magnetic CaTe. The Dirac cone of SnTe remains intact and at the Fermi level. Our results indicate that the SnSe/EuS (111) interface may be promising for further investigation because the topological interface state has a larger gap and a smaller binding energy than those reported for some other materials. 

\begin{acknowledgments}
We thank Di Xiao, Michael Widom, Marek Skowronski, and Alan McGaughey from Carnegie Mellon University and Sergey Frolov from the University of Pittsburgh for helpful discussions. Work on SnTe/CaTe was funded by the National Science Foundation (NSF) through grant OISE-1743717. Work on SnSe/EuS was funded by the Department of Energy through grant DE-SC0019274. This research used resources of the National Energy Research Scientific Computing Center (NERSC), a DOE Office of Science User Facility supported by the Office of Science of the U.S. Department of Energy under contract no. DE-AC02-05CH11231.
\end{acknowledgments}

%

\end{document}